# Understanding Workplace Relatedness Support among Healthcare Professionals: A Four-Layer Model and Implications for Technology Design

Zheyuan Zhang*
Dyson School of Design Engineering
Imperial College London
London, United Kingdom
zheyuan.zhang17@imperial.ac.uk

Dorian Peters
Dyson School of Design Engineering
Imperial College London
London, United Kingdom
Leverhulme Centre for the Future of
Intelligence
University of Cambridge
Cambridge, United Kingdom
d.peters@imperial.ac.uk

Lan Xiao
UCL Interaction Centre
University College London
London, London, United Kingdom
Global Disability Innovation Hub
University College London
London, United Kingdom
l.xiao.22@ucl.ac.uk

Jingjing Sun
Dyson School of Design Engineering
Imperial College London
London, United Kingdom
j.sun23@imperial.ac.uk

Laura Moradbakhti
Digital Mental Health Lab,
Department of Psychiatry,
Psychotherapy and Psychosomatics
RWTH Aachen University
Aachen, Germany
Dyson School of Design Engineering
Imperial College London
London, United Kingdom
l.moradbakhti@imperial.ac.uk

Andrew Hall
CW+
London, United Kingdom
Chelsea and Westminster Hospital
NHS Foundation Trust
London, United Kingdom
andrew.hall@cwplus.org.uk

Rafael A. Calvo
Dyson School of Design Engineering
Imperial College London
London, United Kingdom
r.calvo@ic.ac.uk

## Abstract

Healthcare professionals (HCPs) face increasing occupational stress and burnout. Supporting HCPs' need for relatedness is fundamental to their psychological wellbeing and resilience. However, how technologies could support HCPs' relatedness in the workplace remains less explored. This study incorporated semi-structured interviews (n = 15) and co-design workshops (n = 21) with HCPs working in the UK National Health Service (NHS), to explore their current practices and preferences for workplace relatedness support, and how technology could be utilized to benefit relatedness. Qualitative analysis yielded a four-layer model of HCPs' relatedness need, which includes Informal Interactions, Camaraderie and Bond, Community and Organizational Care, and Shared Identity. Workshops generated eight design concepts (e.g., Playful Encounter, Collocated Action, and Memories and Stories) that operationalize the four relatedness need layers. We conclude by highlighting the theoretical relevance, practical design implications, and the necessity to strengthen relatedness support for HCPs in the era of digitalization and artificial intelligence.

**ACM Reference Format:**
Zheyuan Zhang, Dorian Peters, Lan Xiao, Jingjing Sun, Laura Moradbakhti, Andrew Hall, and Rafael A. Calvo. 2026. Understanding Workplace Relatedness Support among Healthcare Professionals: A Four-Layer Model and Implications for Technology Design. In *Proceedings of the 2026 CHI Conference on Human Factors in Computing Systems (CHI '26), April 13–17, 2026, Barcelona, Spain.* ACM, New York, NY, USA, 21 pages. https://doi.org/10.1145/3772318.3790900

*corresponding author.

# 1 Introduction

Healthcare professionals (HCPs) in the UK and globally face increasing challenges and work-related stress and burnout [23, 67]. Stress, burnout, and poor mental wellbeing can lead to low work engagement, medical errors, and intentions to leave the job [39, 74]. It is crucial for healthcare organizations and systems to provide extra support and resources to alleviate staff burnout, as well as to facilitate positive coping strategies for staff resilience [61]. One of these strategies is facilitating social connection and support among colleagues, which is crucial for HCPs' wellbeing and resilience. Previous studies have consistently shown that social support among colleagues is one of the most critical resources for psychological wellbeing, particularly for frontline HCPs [17, 59, 92]. This aligns closely with the concept of relatedness, a fundamental human psychological need identified in Self-Determination Theory (SDT) [76]. Relatedness refers to the need for meaningful connections and a sense of belonging with others, and is a key driver of work motivation and mental wellbeing [30, 76]. Fostering social connection and peer support among HCPs is crucial for their work motivation, resilience, and wellbeing [2].

Worryingly, over the past decade, particularly during and after COVID-19, HCPs' social connections with colleagues have declined significantly, reducing their social buffers against distress and burnout [49, 79]. The increasing reliance on technology in healthcare, including the growing use of electronic health records (EHR), online learning, and remote communication platforms, has been shown to reduce in-person interactions and diminish social support among HCPs [71, 79].

While technology has often exacerbated disconnection in healthcare settings, it also holds significant potential to facilitate social connections. A substantial body of literature in Human-Computer Interaction (HCI) and Computer-Supported Cooperative Work (CSCW) has focused on products and prototypes that promote social connectedness among individuals. Interventions using computer technologies such as wearable devices, mobile applications, tangible interfaces, and mixed realities have been extensively evaluated to increase sense of connectedness, belonging, and enhance relatedness satisfaction [81, 90]. Hence, digital interventions can be helpful for increasing connections and satisfying relatedness needs for HCPs, given their accessibility, flexibility, and often lightweight design, which could be ideal for HCPs, especially when carefully woven into their routines.

Nonetheless, most existing studies have focused on close relationships (families, lovers, and friends), and research on workplace relationships remains rare, according to a recent review of works from 2010 to 2024 [90]. Even among those studies specifically examining work relationships, the majority have concentrated on general knowledge workers and remote working scenarios [20, 57, 91]. This makes research less applicable to HCPs, who work in a unique clinical context and are less likely to fully adopt remote working styles given the necessities of direct patient care and hands-on team collaboration. Workplace interactions and social support still play an irreplaceable role in fostering relatedness need satisfaction among HCPs [49, 73].

Meanwhile, psychological theories such as SDT, belongingness theory [5], and sense of community theory [62] have established a common understanding of human need for connection and relatedness. However, work that applies these perspectives to HCPs often treats relatedness in broad, generic workplace terms (e.g., "feeling part of a group" or "having connections with others" [47, 85]), offering limited insight into the unique inter-collegial dynamics and organizational contexts of HCPs. This is a critical aspect to be examined, as HCPs are often considered a distinct workforce compared to other occupations [3], and experiencing relatedness support at work contributes significantly to their wellbeing and resilience [47]. Further delineating HCPs' current practices and preferred strategies for relatedness support can help healthcare managers and policymakers better support HCPs on the frontline.

Taken together, some critical gaps remain. More evidence is needed to understand: 1) how HCPs' need for relatedness is supported in healthcare contexts, and 2) how technology could benefit their relatedness support and meaningful connections with colleagues. This study aims to address these gaps with a two-phase study design, using semi-structured interviews and co-design with HCPs of the National Health Service (NHS) in the UK. Our results indicate a four-layer exploratory model that summarizes the current and preferred relatedness support layers among HCPs, which are Informal Interactions, Camaraderie and Bond, Community and Organizational Care, and Shared Identity. This empirically-grounded model extends existing relatedness theories by capturing the depth and breadth of support across personal, team, and organizational levels. We further articulate HCI design opportunities across these layers, offering transferable insights and future directions to support HCPs' need for relatedness amid the digitalization of healthcare systems.

# 2 Related Works

## 2.1 Psychological Theories on Relatedness

Many psychological theories provide valuable insights into relatedness. The belongingness hypothesis [5] argues that the need to form and maintain social relationships is a fundamental human motivation. It points out that human beings desire interpersonal relationships that cannot be satisfied by mere superficial social encounters or interactions with people that one dislikes or has no interest in [5]. In social psychology, Social Identity Theory [46] highlights how identifying with a team, group, or organization (e.g., "being HCPs in one hospital") influences human behavior, values, and actions. Furthermore, at the group and community levels, Sense of Community Theory [62] underscores the roles of membership, influence, shared emotional connection, and the fulfillment of needs in creating a sense of belonging within wards or organizations. Similarly, other theoretical approaches, such as Relational-Cultural Theory [51], emphasize the mutual empathy and empowerment from deep, nurturing relationships, while Social Baseline Theory [22], grounded in neuroscience, demonstrates that having supportive others nearby reduces perceived threat and cognitive load. In this work, these perspectives informed our thinking about different dimensions of relatedness, but they do not constitute an a priori model. Instead, our findings were derived from inductive analysis, while some theories were used to interpret or distinguish them.

In Self-Determination Theory (SDT), relatedness is defined as a basic psychological need to feel meaningfully connected with others [76]. When this need is satisfied, it supports intrinsic motivation, resilience, and wellbeing [37]. More importantly, SDT regards relatedness as a psychological "nutriment" for motivation and wellbeing. By contrast, other theories (e.g., Social Identity Theory [46]) focus primarily on social connections without necessarily making implications for wellbeing. We adopt SDT as the primary theoretical lens as it is one of the most studied and empirically validated psychological theories across various cultures and populations, including HCPs [16, 38]. For example, SDT and basic psychological needs (autonomy, competence, and relatedness) satisfaction are associated with physicians' and nurses' wellbeing [38, 42], work motivation [85], and intention to stay in their positions [44, 47]. Besides, relevant studies have translated insights from SDT into technology design, evaluation, and ethical guidelines, which could have valuable implications for HCI research and practice. Furthermore, SDT also provides widely validated, cross-cultural instruments for need satisfaction and frustration [86], enabling HCI practitioners to empirically assess the impact of technology interventions on relatedness [68].

While many studies have examined SDT among HCPs in medical contexts, relatedness is typically defined in a simplified, global way. For example, many studies among HCPs have utilized SDT-based scales (e.g. Work-related Basic Need Satisfaction scale [84]) on relatedness [38, 41, 47, 63], with items that are more general to overall workplace contexts like "At work, I feel part of a group", and "Some people I work with are close friends of mine". In addition, a growing body of SDT-informed studies, including empirical research that used SDT as a guiding framework [44, 85], and perspective papers advocating improvements in HCPs' need satisfaction [34, 42], typically refer to relatedness in generic terms such as "connection with colleagues," or "more team building activities". Such definitions risk simplification that fails to reveal the nuance and contextualized nature of relatedness needs among HCPs. Our study, in this regard, develops a qualitative and multifactorial account for HCP-relatedness needs, offering a more in-depth and contextualized understanding of how HCPs' relatedness needs are best satisfied at different levels of their work.

## 2.2 HCPs' Workplace Relatedness Support

Relatedness and social support have been widely recognized as key coping mechanisms for HCPs against stress and burnout [6, 61, 73, 79]. Evidence showed that HCPs with adequate social support are more resilient to stress and burnout [79]. This is especially true in the aftermath of the COVID-19 pandemic, where social connectedness has become an indispensable resource for HCPs' wellbeing [6, 13, 59]. One crucial component of this support stems from family and friends [6]. In the meantime, as HCPs spend most of their time within their units and wards, workplace relatedness support is also crucial for HCPs' wellbeing [2, 87].

Compared to many office settings or knowledge worker populations familiar with HCI, HCPs and their workplaces are characterized by unique organizational and medical cultures [53], and shaped by their 24/7 shift systems, high workload, work complexity, and acuity [3, 89]. Besides, HCPs typically work in multi-disciplinary teams with strong professional hierarchies and boundaries (e.g., between doctors, nurses, and support staff) [72, 96]. In such contexts, inter-collegial connections are often limited to handovers, debriefs, and brief interactions during breaks. Furthermore, the professional stigma and norms can discourage HCPs from disclosing topics like distress and burnout [19, 29]. Hence, it is crucial to be mindful of the unique context and dynamics when developing relatedness interventions for HCP.

Several strategies have been studied to bolster connections among HCPs in the workplace, including regular peer-support sessions [14], mentoring or peer-buddy programs [79], provision of shared social spaces [83], promotion of mutual support and open communication [82, 87], and improved debriefing structures and rota (shift schedules) management [2]. For example, Uys et al. [83] highlighted the significance of having common spaces for HCPs' resilience and wellbeing. Meanwhile, peer-support programs, in which HCPs engage in open discussions about challenging situations, have also been explored across healthcare populations globally [14, 48]. In summary, relatedness support strategies identified by previous studies often operate at the organizational level. Digital approaches could be more easily scaled and personalized, and some of the prior work in this space is described below.

## 2.3 Technology and HCI for Relatedness

Numerous studies in HCI have examined the design and implementation of technologies that support relatedness. In a recent literature review, Stepanova et al. identified some common strategies that foster "a genuine feeling of connection" using technology-mediated systems, such as affective self-disclosure, reflection on unity, shared embodied experience, and interpersonal distance [81]. Wenhart et al. proposed nine design strategies commonly applied in technologies designed to foster relatedness, which are awareness, expressivity, physicalness, gift-giving, joint action, memories, genuine conversation, acts of care, and ritual formation [90]. Although prior work offers valuable insights, most studies focus on close relationships, such as romantic and intimate relationships [43], family relationships (e.g., parent-child relationships) [78], and friendships [7, 90]. Work relationships, and more specifically those among HCPs, remain less addressed [90].

Among the limited number of studies on workplace relatedness technologies, Shen and Kelly designed a smart coaster that promotes awareness of colleagues' availability by sensing if their cups are on the coaster, exploring how lightweight and ambient information sharing could influence interactions and connections in the workplace [77]. Chow et al. explored how computer-mediated self-disclosure could be designed for workplace scenarios [21]. Gallacher et al. focused on sparking conversations and interactions among colleagues using squeezable balls and interactive visualizations in the office [31]. Meanwhile, a growing number of studies have shifted direction from the physical workplace to hybrid/home office scenarios, fostering awareness of remote colleagues [57, 88, 91] and enhancing interactions and meaningful conversations at a distance [75].

Within healthcare, although relatedness technology is less studied, many studies on digital interventions for the general wellbeing of HCPs offer some insights. Several recent reviews of digital health interventions for HCPs [1, 45, 58, 94] have identified issues such as high dropout rates. The key factors that may lead to dropout and low engagement with digital interventions among HCPs include a lack of structural and managerial backing, stigma and privacy concerns, lack of tailored content, and difficulty fitting into clinical routines [1, 69]. Interestingly, although focusing on general stress and wellbeing, one systematic review of digital interventions [58] hypothesized that the short-term benefits of guided digital interventions may stem from the formation of a therapeutic alliance between HCPs and facilitators, suggesting that relatedness support is vital for the engagement and benefits of wellbeing tools for HCPs. Similarly, another recent study on HCPs' engagement with digital wellbeing tools discovered that a lack of human element and social support was one of the key factors that hindered HCPs' motivation to engage [95]. This indicates the importance of supporting relatedness with digital tools, which this study will explore.

## 3 Methods

To explore HCPs' current practices and preferences for relatedness support and identify potential technology designs that can facilitate relatedness support, we adopted a two-phase study design, including semi-structured interviews and co-design workshops. Insights from the interviews informed the design and analysis of the subsequent workshops.

### 3.1 Recruitment and samples

A volunteering sampling method was used for both interviews and workshops. Participant recruitment took place at Chelsea and Westminster Hospital, a public NHS hospital. Recruitment messages for both the interviews and co-design workshops were distributed via A4 posters displayed within the hospital's communal areas and staff rooms, and through digital messages shared in hospital community groups and messaging channels. To be eligible for both studies, participants had to be over 18 years old and employed as staff members (either clinical or administrative) in the hospital when the studies took place.

31 HCPs responded to the recruitment for interviews. However, four withdrew due to workload, 10 didn't reply to follow-up emails, and two didn't finish the interview due to personal and work emergencies. Their feedback was excluded as insufficient after being reviewed by the research team. Eventually, 15 HCPs finished the interviews, among them 4 were doctors, 7 were nurses, and 4 were administrative staff. Participants were anonymized and referred to by their unique participant ID, with doctors as D01-D04, nurses as N01-N07, and administrative staff as A01-A04. Detailed participant demographics are listed in Table 1.

For the co-design workshops, 67 respondents replied to the recruitment message. Eventually, 21 participants registered and were divided into five separate workshops to meet their shared availability. Workshops were held in four hospital communal areas or meeting rooms located near participants' departments. Two participants left early, but their input and ideations were retained in the dataset. Similarly, participants were anonymized and referred to by their participant ID, with P01-P21 for workshop participants. The co-design workshop and the interview recruitment happened separately. Because identifiers were removed before analysis, we cannot determine with certainty whether any individuals participated in both phases. To minimize potential overlap, recruitment for the workshops was targeted at a different time period (6 months later) from the interviews.

### 3.2 Study Design and Data Collection

*3.2.1 Semi-structured Interviews.* Interviews were conducted with HCPs, including doctors, nurses, and administrative staff. The topics included mental health, coping methods, and personal strategies and preferences for both relatedness and social support. The interviews are part of a larger project on technology and HCPs' wellbeing, hence the interview guide is inclined towards stress and coping while investigating relatedness. Within this broader focus, we included specific questions about perceptions of relatedness, experiences of inter-collegial connections, experiences of feeling connected or isolated at work, and preferences for relatedness and social support. Our intention was to understand how the need for relatedness is satisfied in the everyday work of HCPs and to explore its benefits for wellbeing in a real-life context. For the interview phase, participants were not briefed on SDT, and the interview guide was not structured around SDT constructs. Detailed interview guidelines are listed in Appendix A. Interviews were conducted online via Microsoft Teams and Zoom and lasted 25 to 50 minutes. All interviews were conducted by the first author and were voice-recorded with participants' prior consent.

*3.2.2 Co-design Workshops.* Co-design workshops were divided into three stages. First, participants were briefed about SDT and its three basic psychological needs, including relatedness. Participants were also introduced to case studies and related technologies that have been explored in HCI and design to support wellbeing and relatedness needs (see introductory materials in Appendix B). Participants were then asked to propose ideas for designing technologies to meet HCPs' need for relatedness in their working context, without specific restrictions on the type of technology or implementation settings. Participants were advised to sketch or write down their ideas using the provided design templates (see Figure 1). Finally, participants were asked to share their ideas and discuss in groups. Each workshop was designed to last around 1 hour, and eventually ranged from 31 to 53 minutes, due to participants' limited availability. All workshops were voice-recorded with participants' consent. The drawings and writings of participants were documented. Figure 1 shows the design template provided to participants (left) and examples of participants' ideations (right).

### 3.3 Data analysis

We adopted an inductive reflexive thematic analysis (TA) approach [9]. In reflexive TA, researcher subjectivity is treated as an analytic resource rather than a source of bias [10]. Hence, we did not run parallel independent coding or quantitatively measure the intercoder agreement, but carried out reflective sessions among the researchers, as recommended by Braun & Clark [10]. The first author, who led the coding and conducted the TA, has been working on design and user research for over 10 years and on research projects with HCPs and in healthcare contexts for over 6 years. The other two researchers who participated in the reflection and formation of themes are both specialized in wellbeing, design, and HCI. NVivo for Mac (version 12; QSR International Pty Ltd) was used to build up codes and organize the node hierarchies, but not to determine analytic decisions on the final themes.

**Table 1: Participant ID and demographics [a]**

| Participant ID (Interviews) | Gender | Department | Participant ID (Workshops) | Gender | Department |
|---|---|---|---|---|---|
| D01 | Male | Outpatient | P1 | Male | Radiology |
| D02 | Female | Psychiatry | P2 | Female | Radiology |
| D03 | Female | NICU | P3 | Female | Therapy |
| D04 | Female | NICU | P4 | Female | Medical Education |
| N01 | Female | Pediatrics | P5 | Male | IT |
| N02 | Female | AAU | P6 | Female | Outpatient |
| N03 | Female | Day Surgery | P7 | Female | Outpatient |
| N04 | Male | Day Surgery | P8 | Male | HIV |
| N05 | Male | A&E | P9 | Male | ICU |
| N06 | Female | ICU | P10 | Male | NICU |
| N07 | Female | ICU | P11 | Female | Therapy |
| A01 | Male | Department Head | P12 | Male | ICU |
| A02 | Female | A&E | P13 | Female | ICU |
| A03 | Male | Therapy | P14 | Female | A&E |
| A04 | Female | AAU | P15 | Female | A&E |
| | | | P16 | Male | AAU |
| | | | P17 | Female | ICU |
| | | | P18 | Female | ICU |
| | | | P19 | Female | NICU |
| | | | P20 | Male | AAU |
| | | | P21 | Male | AAU |

[a] Abbreviations: NICU = Neonatal Intensive Care Unit; AAU = Acute Assessment Unit; ICU = Intensive Care Unit; A&E = Accident and Emergency

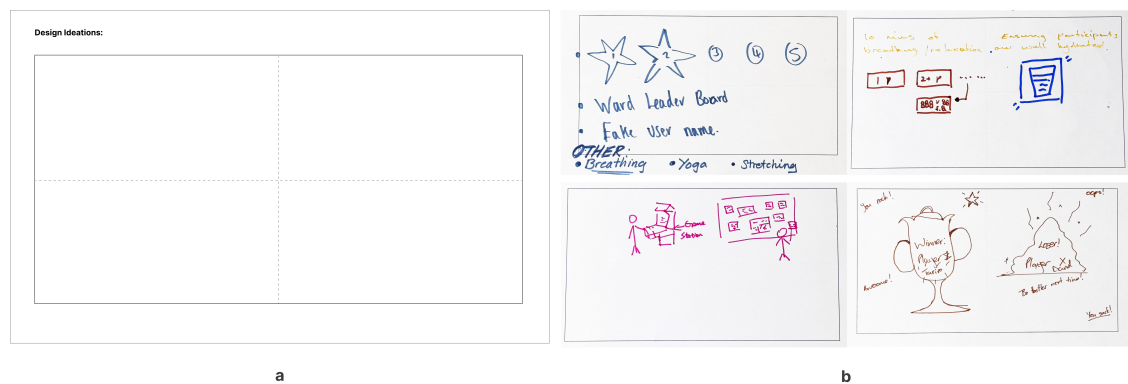

**Figure 1: Design template provided to participants (a) and examples of participant ideations (b)**

For the interviews, qualitative data were extracted from the audio transcription of each interview. The first author first went through the coding process, including initial data familiarization, repeated reading, and inductive coding of the whole dataset, with reflective notes. The first author then performed an initial grouping of similar codes into early patterns, which were then discussed and checked with the other two researchers. This allowed us to identify recurring topics (not themes) such as "Venting to Colleagues", "Positive Affirmation", and "Traumatic Memories of COVID-19". We then returned to the dataset to review, revise, merge, or rename topics, and to develop sub-topics under each main topic. After the coding process, the first author began generating themes from the coding results while reflectively reviewing the codes, code prevalences, and detailed participant quotes. Critical discussions among the researchers were held to examine the initial themes and test alternative interpretations reflectively. The resulting themes, as well as the coding summary, participant quotes, and reflective notes, were then presented to all co-authors to allow collaborative reflections and discussions before the themes were finalized.

During theme generation, the first author proposed further grouping the themes, which was discussed with the other two researchers who participated in the reflexive TA. This grouping is based on similarities among themes and is therefore a topic-summary structure rather than an additional interpretive theme [10]. This grouping eventually formed the relatedness "layers" in the results. The layers highlight the common aspects of relatedness themes and help demonstrate the varying depth and breadth of HCPs' relatedness needs. As the layers are a representational framework for organizing related themes, while the analytic claims still rest within the underlying themes, we argue that grouping the themes as "layers" is not a methodologically incongruent practice that cannot be justified, based on Braun and Clark's guidelines on reflexive TA [10, 11].

For the workshops, data were extracted from audio transcriptions, participant handwritings, and field notes taken by the first author. The first author coded materials and drafted interpretive notes capturing recurrent design topics (e.g., collocated, shared embodied, and embodied activities). Themes were derived from the analysis of the transcriptions, iteratively refined through critical interpretation and analytic dialogue among the three researchers, and checked against the original data to ensure they represented participants' feedback. Meanwhile, each hand-written, sketched, or verbal idea was treated as a design input, and the researchers collaboratively and inductively cluster similar ideas into higher-level design concepts (e.g., Playful Encounter, Collocated Action), which were then mapped onto the interview-derived relatedness layers.

## 3.4 Study Ethics

Studies were approved by the Research Integrity and Ethics Committee of Imperial College London (Reference number: 22IC7803). Ethics approval was granted for the studies in Chelsea and Westminster Hospital by the Health Research Authority (HRA) of the NHS in the UK (reference number: 316935). Informed consent was obtained from participants before the interviews and workshops. Participants were fully informed about the study's objectives, procedures, and potential risks. All data collected in the study was anonymized, and no personally identifiable information was kept. Each participant received a minimum compensation (£10 or equivalent) for their time in the interviews and workshops.

## 4 Results

We present the findings from the interviews (Section 4.1) and workshops (Section 4.2) below. In Section 4.1, we describe how insights from the interviews informed the development of a four-layer model of workplace relatedness among HCPs, supported by our rationale and illustrative participant quotes. In Section 4.2, we introduce the workshop findings through design concepts that align with the framework established from the interview results, highlighting participants' various ideas on technology supporting relatedness.

### 4.1 Layers of Workplace Relatedness Support Strategies among HCPs

As illustrated by Figure 2, the research team grouped interview-derived themes based on their common aspects, into 4 categories, or "layers." These four layers summarize and highlight the variety in the scope and depth of relatedness support strategies among HCPs.

Specifically, the first layer, Informal Interactions, highlighted the immediate, in-the-moment interactions with colleagues that offer HCPs relatedness support. The second layer, Camaraderie and Bond, highlighted the overtime ties and deeper connections and trust among HCPs within their teams and wards. The third layer, Community and Organizational Care, stressed the inter-departmental or organizational support they received, often provided by the hospital where they work, or by a specialized team or organization within a region. The fourth layer, Shared Identity, highlighted the collective, profession-wide social connections and sense of "us" that is both shaped by the previous three layers, but also by the broader healthcare system (in our study, the UK's NHS system), its culture, values, and social-political contexts.

Below, we will discuss the layers with participants' original quotes.

#### 4.1.1 Layer 1: Informal Interaction. 
The first layer summarizes two themes that underscore the significance of everyday social interactions among colleagues, and the crucial conditions that make these interactions beneficial and contributive to HCPs' relatedness. Specifically, HCPs described these social interactions as restorative, informal, and spontaneous, emphasizing their unplanned and in-the-moment nature.

##### 4.1.1.1 "Everyday relatedness" is derived from short interactions within tight schedules
Most participants strongly emphasized the importance of daily interactions and communications with colleagues and their impact on their feelings of connection and relatedness. Participants described them as contributing to their wellbeing and coping with occupational distress. Such interactions included small talk, shared laughter, and personal conversations, or occasional light-hearted moments with colleagues. These interactions were valued by participants, as D03 explained:

> "We have often these moments where we have chats amongst each other. And that might be about patients, or just, in general, how we are getting on with life or things. And I quite like these basic interactions" (D03)

Meanwhile, participants noted that these interactions were often confined to immediate work settings or their tight schedules and routines. The opportunities to extend these daily social interactions into longer breaks or out-of-work contexts are limited, due to their compact schedule and clinical demands, as illustrated by N01:

> "Most of our interactions happen close by within the ward … by the time you get changed, and get to say

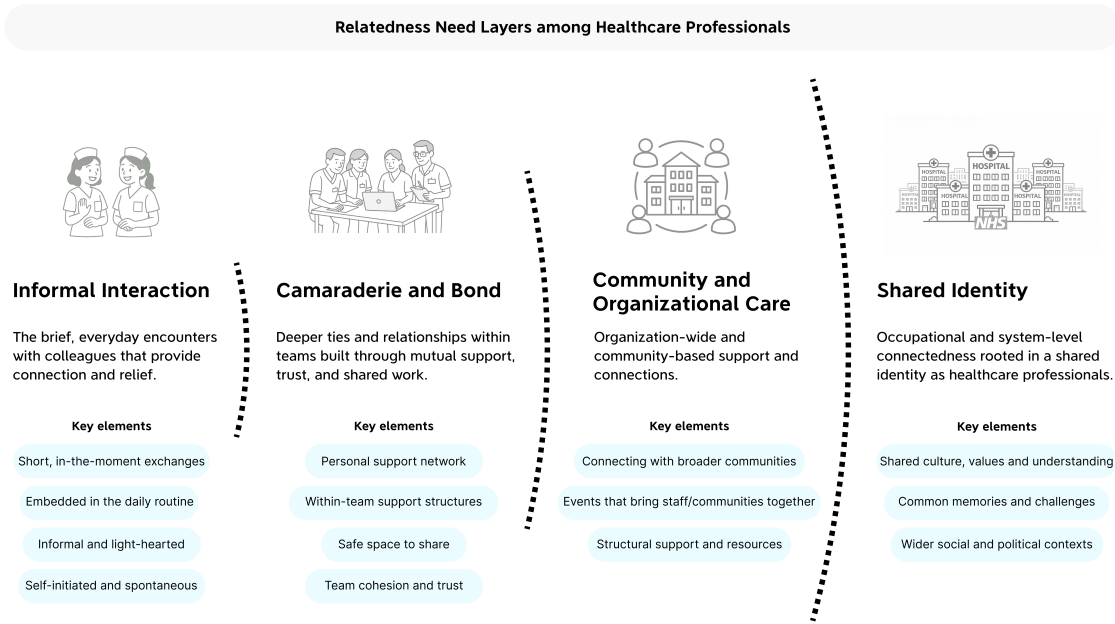

**Figure 2: Workplace Relatedness Need Layers among HCPs**

Costa[1] or anything else, that's half the break done. So people generally stay in the canteen, or in the coffee room within the department" (N01)

Taken together, these accounts suggest that everyday relatedness is built in short moments that fit inside tight schedules. In our interpretation, it is indeed these day-to-day social interactions that potentially embed relatedness and social support into HCPs' work routine, providing relational boosts and wellbeing buffers.

*4.1.1.2 Restorative interactions are often informal, self-initiated, and spontaneous*
One key characteristic of the daily interactions that are cherished by HCPs is that they are often self-initiated, spontaneous, and informal. Although pre-planned activities, like structured wellbeing sessions, peer support groups, and special staff events, offer additional relatedness support, participants stressed the unique and indispensable value of the non-facilitated and informal communications. Because they were unscheduled and unmediated, these interactions felt more natural and less stressful, allowing HCPs to speak up without "making a thing of it" (N04). For many, these spontaneous and incidental interactions among colleagues provide them with an adequate social buffer against adversities at work. N05's reflections exemplified this:

> "For me … much of the peer support happens probably very informally, because colleagues will talk to each other and say, I'm struggling, I'm having a hard time. And people will notice that colleagues are struggling and having a hard time." (N05)

Some participants further argued that facilitating restorative social interactions is challenging from a top-down approach. Instead, they tend to emerge organically, independently, and spontaneously from within the staff dyads or small groups. For example, A03 and D02 both described low attendance rates and a lack of staff motivation in the formal social support programs. As A03 noted:

> "It's the informal and self-initiated communications that play a bigger role, I'd say. And what I see is often when you try to put formal programs around, is where you find people may not come to them." (A03)

D02 agreed with this and stated that a major part of social connections among HCPs occurred briefly and naturally. Hence, it can be valuable to explore how to naturally create more opportunities for informal and spontaneous interactions among HCPs.

**4.1.2 Layer 2: Camaraderie and Bond.** The second layer highlighted the deeper bond and a sense of camaraderie that HCPs formed with their colleagues within a team or a ward, through mutual support and teamwork over time. This represents a deeper relationship among HCPs compared to ordinary day-to-day interactions. The two themes in this layer illustrated the importance of personal support networks, teamwork, and trust, which constituted this deeper bond among HCPs.

*4.1.2.1 Everyone needs a personal support network within the team*
Participants highly valued the bonds within their teams and wards, which had been established through close collaboration and reciprocal caring and compassion. Many mentioned that they relied on their own go-to support network within their team, which they turn to whenever they are feeling unwell or overly stressed at work. Building such a personal support network was regarded as especially beneficial to their daily practice and overall wellbeing. In practice, this support is frequently derived from a dependable colleague (as mentioned by A02) or a small circle (for N02):

> "There is a very good support network among us in Neonate. So we have a palliative care, complex care nurse, who is really, really good at organizing, debriefing, and even a 1 to 1 pull-out-of-the-room. Let's go for a coffee because you don't look well ...she's like a person who makes the whole team connect ..." (A02)

> "Definitely when I was a junior member of staff, the team around me, not my senior team, like my peer junior nurses were my support bubble." (N02)

---
[1]A cafe chain in the UK and worldwide

Besides these individual support networks, many teams also have their own small-scale support structures that colleagues trust in, in which social support is often provided by supervision or ad-hoc debriefs. A04, for example, valued their own support structure within their teams:

> "I've got good support structures in the team I work with. So there's a supervision structure that we have. We have regular supervision with my senior and also quite informal debriefings with the other disciplines we work with." (A04)

Through our interpretation and reflections on participant input, it became evident that these personal ties and within-team structures created a first line of support for HCPs. It is through these entrusted support networks and team structures that participants felt cared for and deeply connected with their colleagues.

#### 4.1.2.2 A safe environment is crucial for deeper connections and teamwork

Besides highlighting the significance of closer bonds and connections, participants also stressed that such deeper connections depended on a psychologically safe environment where it feels acceptable to seek and offer support. In other words, HCPs' bonds and mutual support were enhanced by an environment in which they felt able to share their concerns and distress without being judged or harming their relationships with others. N03, for example, reflected on the team dynamics in their own unit, which, to her, provided a safe and dependable support:

> "To say, in this case, I had a family that was quite difficult. Or I had a patient's death who it's just kind of stuck with me, I can just vent or ask for help within our unit ... So it's a safe space ... to just release and be able to say anything that was bothering you." (N03)

Some participants in senior roles (such as N03 and D04) described intentionally influencing the climate to create a safe environment for colleagues, by making themselves more approachable, inviting concerns upward, and effectively protecting junior staff members to foster team cohesion:

> "So I always say to my direct report, if you want to vent, fine, vent to me, and they would never vent downwards. And we sort of have that philosophy with our matrons ... because actually, the people who report to you don't need to know your problem." (N03)

> "The juniors in our unit have great teamwork ... From one side, I do encourage them to support each other,

because they know what they are going through. At the same time, now that I'm a senior staff member, I want to support my team in a way that I didn't feel supported at the time." (D04)

According to our participants, such a psychologically safe environment enabled more trusting teamwork and mutual support. Some mentioned relying on their colleagues when facing complexities and an increased workload, which has made things much easier through teamwork and the team cohesion fostered over time. This factor is especially relevant for HCPs, whose work is inherently interdependent. N06 mentioned her experience with a cohesive and supportive team:

> "... what we do is you divvy the work as a team. And if one patient is taking you longer ... you can always go back to the team and say, can somebody help me? ... I see that the team was really good in taking on the other patients that I was supposed to see, so I was able to focus on that one ... and this reduced my stress a lot." (N06)

#### 4.1.3 Layer 3: Community and Organizational Care.
This layer focuses on the relatedness support among HCPs that comes from their sense of connection and belongingness to broader communities and organizations. Where the previous layer centered on interactions and bonds within immediate teams, wards, and personal circles, this layer reflects how connection extends to cross-unit, organization-level communities and supports. Two themes in this layer reflected participants' perceived value of being involved in communities, and the importance of organization-based structural support, which provided them with the confidence and reassurance that they were not alone, and that they would be understood and supported by their community.

#### 4.1.3.1 Connection to broader communities buffers distress and isolation

Participants highlighted the significance of building connections with a broader community, gaining extra social and relatedness support beyond their own team. According to participants, these broader community ties acted as an antidote to their feelings of being alone and isolated. For example, A04 and A03 highlighted their preferences for more community participation, which has the value of connecting HCPs with broader cohorts beyond their own circles:

> "So even something like a bike group across the Trust[2], or you know, like a social group ... or a community event or something ... Rather than kind of, a tranquil place to sit for lunch or, like a pod to lie in, I'd prefer something that makes me feel connected to a bigger group ..." (A04)

> "I always have this idea ... if you find people within our hospital who were close where they live, then have, say, ten in each area, like southwest London, west, north, and northeast, it allows people to then use that kind of community benefits of your organization." (A03)

---

[2]In the UK, an NHS Trust is a publicly funded organization that manages one or more hospitals and services

For many of our participants, more initiatives that made it easy to find "people like me" across the organization were highly valued. This, from our analysis, may reflect the fact that they work in a large NHS hospital, and hence making this large, complex organization feel more relational and personally connected, could further support their relatedness needs.

Furthermore, a feeling of connection and active involvement with broader medical or healthcare communities is beneficial for their relatedness needs as well as mental wellbeing. For example, N07 reflected on the value of community support and the sense of "you are not alone", which buffers against work-related distress and anxiety:

> "Knowing that you're not alone is quite important. Because actually, what if I'm a first-time matron and I'm in my role, I'm really overwhelmed. I think I'm the only person … I'm like, it must be me, I just might not be good at my job. But if you are in like an active community … other people would be like, actually I'm in the same situation..." (N07)

Engaging in community activities and support programs is especially important to mitigate a sense of loneliness and isolation. N01 brought up his personal experience with organizational peer-support groups and their contribution to the sense of community and togetherness:

> "So sitting with people … shared thoughts and have experienced similar things … There's something really important in that, cause then you realize and recognize I'm not alone. I'm not the only one who feels this, even if you didn't contribute to that session." (N01)

#### 4.1.3.2 Relatedness benefits of organizational care and structural support

Structural and organizational support were also valued approaches to develop connections with a wider community, as described by participants. N02, for example, recalled the events organized by the hospital, which were valuable to her in terms of fostering a strong sense of community and togetherness: "I like when the hospital has like some weeklong events, I think things like that are really good for a sense of community." (N02). She continued to praise such initiatives of the hospital, emphasizing how such events bring staff together and foster a feeling of belongingness:

> "I feel like I'm part of the community, then I have a motivation to work better to get on better with people to work as part of a team. I think when it comes to wellbeing, anything involved with creating a community, atmosphere is something I would benefit from." (N02)

Similarly, A01 reflected on the benefits of organizational events that enabled shared experiences among staff:

> "Yesterday was the 30th birthday of the hospital. There were lots of celebrations … so many members of staff dancing, having fun, there was music that playing … It wasn't a formal stress-reduction session, but we all felt relieved … And I think there is something really important about the shared experience that bring you together as an organization." (A01)

Taken together, it is evident that the organizational events that create shared experiences are often regarded as boosting relatedness and wellbeing. In such events, HCPs temporarily suspend routine roles, create shared positive memories, and experience themselves as part of the organization rather than isolated units.

Meanwhile, other participants recognized the value of structural support and programs, and regarded them as crucial relatedness and wellbeing resources. N04, for example, expressed his pride and gratitude for the variety of support programs within the organization:

> "And we've got good resources … from an organizational perspective, we've got the staff groups here around to do Mindfulness, yoga, we've got the mental health first aiders, the wellbeing champions. We've got psychologists who could support ..." (N04)

We observe that for many HCPs, structural resources do more than offering individual coping tools. They also represent that the organization is willing to support and care for its staff. This, in turn, reinforces HCPs' connection with the broader community and organization and could benefit their wellbeing and work engagement.

#### 4.1.4 Layer 4: Shared Identity.
The final layer we identified was the importance of a shared identity to a sense of relatedness among HCPs. While the first three layers describe how relatedness is enacted through everyday interactions and support structures, the final layer reflects a more cognitive aspect of relatedness among HCPs, shaped not only by workplace experiences but also by the wider culture, societal, and political contexts.

In this study, participants noted that a shared understanding of their roles and a common sense of responsibility and commitment could bind them together. Meanwhile, shared memories and challenges further shape this identity and consolidate a feeling of togetherness. This shared identity transcends departments and organizations, representing a unified concept of being HCPs in the NHS, which connects them and provides an additional layer of relatedness need satisfaction.

#### 4.1.4.1 Shared responsibility, challenges, and COVID-19 memories consolidate a sense of "us"
Participants highlighted the responsibility of their roles, and their identification with the roles and the accompanying responsibilities formed a crucial aspect of their shared identity as HCPs. Participants like D03 mentioned that such shared responsibility, values, and commitment are the foundation of their trust and relationship, with the expectation that colleagues uphold common purposes and standards of care. N06's reflection also echoed with this aspect:

> "They want to support them … And that's the whole idea that you're putting patients first. And that's expected among people, no matter if you know them or not." (N06)

Another aspect of the shared identity that became salient during the interview was the common challenges and memories HCPs shared. Most participants involved in the interview went through the COVID-19 pandemic, which left them with memories filled with stress, anxiety, and even struggles. Nonetheless, the common challenges and memories appeared to shape and consolidate their shared identity as HCPs working within a public healthcare system such as the NHS. According to our participants, the pandemic created a strong sense of "us" and "only-we-know" experience that bound colleagues together across roles and organizations.

> "You had staff members here and across the NHS ... who were nervous and anxious and even sometimes plain scared about the situation. It was the unknown, it was the unprecedented, what we were through ... it's the thing that only we understand and feel it in this way, the people who have been through this ..." (N05)

Similar to N05's reflection, many participants believed the challenges and difficulties were shared across the NHS and displayed compassion for frontline colleagues beyond their own organization. This sense of sharing the past, current, and future challenges could consolidate their shared identity.

*4.1.4.2 The importance of feeling appreciated and valued by society and the NHS system*
With their sense of responsibility and commitment, many noted that it is rewarding to see their work being appreciated and recognized. Besides micro-recognitions within the team or more formal acknowledgements from the organization, public recognition is also important, which strengthens their identity as HCPs serving the public good. Such rewards and positive reinforcement as a collective identity are also beneficial.

D04, for example, brought up the events of "Clapping for the NHS" during the pandemic, regarding it as crucial for empowering frontline staff:

> "It was quite a warm, encouraging activity that even if it does not directly help you out, it still gave you that power and mental strength to hang in there, for everyone." (D04).

In the meantime, participants were clear that recognition alone was not enough. As HCPs working in the public health system, which faces multiple structural problems, a few participants expressed a strong desire for more systematic shifts and improvements to support staff. D01 described issues in the Covid aftermath, such as short-staffing, and increasing work complexity, and linked relevant policy and funding changes to HCPs' feeling genuinely valued:

> "(After Covid) We then move into a recovery phase within the NHS where we're trying to kind of catch up on activity that was canceled, asking people to work harder and longer and faster ... And I think people were really struggling ... because they just had to keep working, keep going ... I do think there is a need for more policy change and more resources to the NHS ... it is critical that people feel appreciated for all their sacrifices" (D01)

These suggest that the HCPs' shared identity is multi-faceted and shaped not only by their shared goals and commitment, but also by public recognition and concrete policy- and system-level support.

## 4.2 How Technology Can Support Meaningful Connection among HCPs

In the co-design workshops, participants were invited to generate their own ideas that use technology to facilitate relatedness. Participant ideas were analyzed and categorized into eight concepts by the research team, which span across the four relatedness need layers, as shown in Figure 3. We will go through the design categories (or concepts) below based on their shared focus and goals in supporting relatedness. Table 2 further illustrates the categories of ideas, their respective need layers, and exemplary ideas.

*4.2.1 Technology could facilitate and improve interactions among colleagues.* Throughout the co-design workshops, creating more interaction among colleagues was highly valued by participants. For example, **Playful Encounter** was a concept endorsed by many participants, including ideas that use digital tools to facilitate interactions and icebreakers among HCPs in a common space. P7, a junior doctor, favored this idea and reflected on her experience that "... it's a shame that when we do rotations, I didn't have a chance to meet and know many of the staff there". She went on to showcase her design for a shared desktop speaker for the staff room, which could prompt social cues and questions to facilitate icebreaking and mitigate embarrassment and shyness.

Furthermore, the concept highlighted playfulness, emphasizing that social interactions among HCPs could be initiated in a harmless, playful context. Participants like P7, P8, P13, and P19-21 all mentioned ideas that involve playful competition among colleagues to foster interactions and mutual bonds. P13 demonstrated her idea on competitive games:

> "I would love to have some quick, relaxing games that you could play with others ... Many doctors in our team, I think we'd like aspects of competition and game, if the purpose is for, like, team building or whatever, we would love to see some competition ..." (P13)

Similarly, the concept of **Collocated Action** also appeared in many ideas generated by participants. For example, P1, P9, P12 and P16 brought up the idea of having VR, projections or screens for teams and groups to engage with collocated activities like yoga, breathing exercises, or just simple "digital retreats". Sharing activities and experiences could be ideal opportunities for HCPs to interact and form deeper bonds, as recommended by P1:

> "I already quite enjoyed the current mindfulness session in the hospital Studio ... If it's a digital thing, then I reckon we could use it more often and get to know more people in the Trust" (P1)

The aspect of collocating was crucial for many participants. Being able to communicate, relax, and share their thoughts face-to-face was regarded as fostering a deeper sense of connection and enjoyment.

*4.2.2 Technology could support team connection and cohesion.* Similarly, participants were enthusiastic about exploring how technology could support team cohesion by offering opportunities for team building and interpersonal bonding. The concept **Team-up Activities** summarizes participants' ideas on how technologies can be designed to provide opportunities for team building and bonding. It fosters greater interaction among colleagues while creating opportunities for deeper bonds and camaraderie. P4, P10-12, P15, P16, and P21 provided ideas on unique and time-efficient team-building opportunities with technology. Opportunities for team collaboration or competition in a playful context (P4), inheriting or transferring team identity into a virtual world (P10 and P21), and shared digital artifacts (like a digital garden) for teams (P15 and P16) are some representative ideas that showcase their preferences. P11, for example, introduced her idea:

> "I'd like to see teams engaging in same activities … for instance if there's like, a game with the option to sign in like as yourself or sign in like as a team. So if we could be like, as a therapy team, then we could … have built up points for our team and I guess that would be pretty fun." (P11)

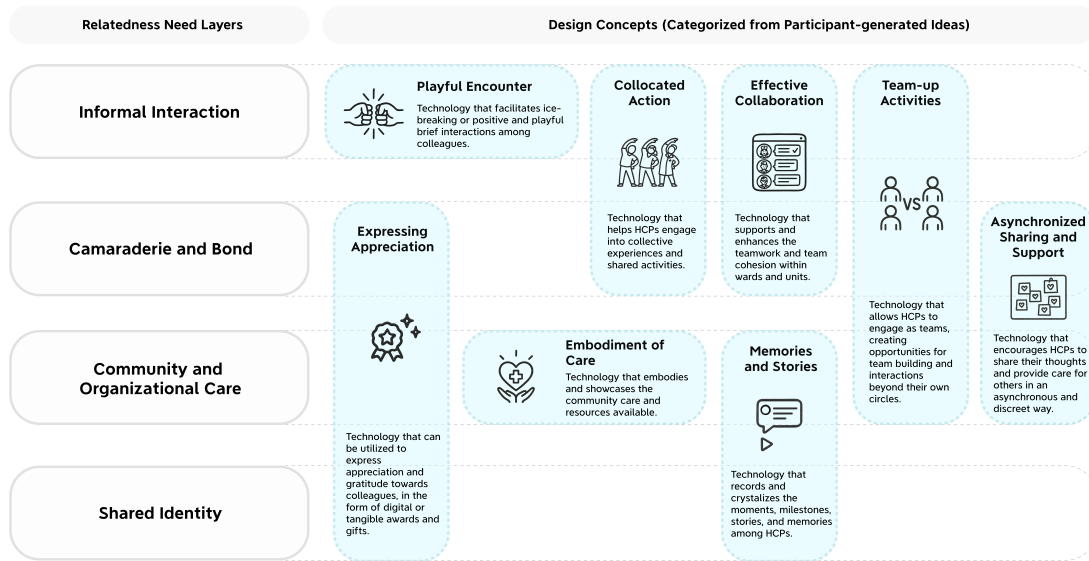

**Figure 3: Design concepts categorized from participant-generated ideas and mapped on the relatedness need layers**

On the other hand, participants envisioned how technology can benefit cross-disciplinary team collaboration, which is key to team cohesion and efficiency. The concept of **Effective Collaboration** represented their ideas on technology enhancing teamwork and communication. Ideas like an updated EHR system where HCPs can see an AI-based wrap-up of similar cases by other colleagues (P2), a chatbot that helps summarize daily debrief for effective teamwork (P7 and P8), and a VR training for empathetic communication within teams (P10), all focused on the use of technology to boost teamwork and collaboration.

*4.2.3 Technology could convey care and understanding on a community level.* The possibility of using technology to convey and spread care and understanding to colleagues was also highlighted by participants. To many, new technology can be a platform for care and empathy, providing an extra buffer against stress and burnout at the community level. For example, participants reflected on the importance of team- and community-based care and support and generated ideas that enable more discrete conversations, sharing, and disclosure within or beyond their team. For example, the concept of **Asynchronized Sharing and Support** summarized participant ideas to create a digital space for disclosure and reciprocal support. P1, P3, P4, and P14 all proposed similar ideas for a digital app or a screen-based installation in the ward, on which staff can disclose their needs, challenges, and concerns, while others can offer support. P14 imagined a virtual chatroom where everyone speaks freely in a safe and anonymized virtual space:

> "And I wonder whether some kind of digital space for group connection, but without you physically having to sit in a group in front of people, where people can kind of share experiences, stories, and other people can comment on that … I could go on as an avatar or something and just say my name's John or Lee …" (P14)

Such digital intervention could help HCPs access care and support at the community level, potentially improving their sense of connection and belonging. Similarly, another concept, **Embodiment of Care,** also probes into how to provide that extra community support, but on an organizational level. Participants spoke highly of the redesign in some communal areas of the hospital, where the additional elements were particularly meaningful in embodying the organization's care and support. P5 and P6, for example, mentioned their experience with the sleep pods introduced to the communal areas during COVID-19 and praised them for embodying care and understanding.

> "I like the sleep pods where you have screens for music, mindfulness programs. Even if I don't use it that often, it's nice to have something that makes you feel that people are doing something, people are listening to you, and trying to care." (P5)

Therefore, they envisioned similar interventions that use tangible and embodied interfaces to facilitate effective relaxation and stress relief sessions. However, the value lies not only in the wellbeing activities but also in the technology itself, as an embodiment of the organization's care and support.

*4.2.4 Technology that helps consolidate their identity and commitment.* Several participants also proposed technological interventions to strengthen the feeling of being recognized and appreciated. The concept **Expressing Appreciation** represents their ideas around using digital artifacts and signage to spread recognition and a positive message of staff members' contributions and devotion. Such appreciation can take the form of virtual badges (P7 and P10), digital posters (P3), or tangible collectibles (P17). More specifically, P14 reflected on the current digital signage in the hospital, and proposed a "Digital Wall of Gratitude" idea:

> "You know the comms teams put up screens with the posters showing our photos and the stop violence on staff messages ... I was thinking, can this be a community wall of fame, or wall of gratitude, and people can all contribute with their thankful messages and stories to each other, and it keeps on updating every day." (P14)

Participants also recognized the potential to use technology for documenting key moments within or across organizations, which forms the concept of **Memories and Stories**. P3 and P5, who worked in the frontline during the pandemic, recalled their encounter and adoption of an iPad in their ward office. It was initially designed and implemented for a sound system control, but was adopted by the team as a selfie station where they record key moments in the ward:

> "We used to have an iPad on our office wall ... but in the end people started to take selfies. I still remember one group selfie we took after we received a long-expected parcel for the office ... that photo is still in my phone and we really would like that 'photo booth' back ... I guess we can even link different photo stations and spread the stories across the NHS" (P3)

Their story and ideas revealed the potential of technology as a medium for documenting and crystallizing special moments and memories within a team and an organization. It could contribute to HCPs' team morale, as well as a symbol of their common challenges, shared identity, and commitment.

## 5 Discussion

In this study, we have identified four salient "layers" that represent the interview themes and support HCPs' relatedness in the workplace. These are: Informal Interaction, Camaraderie and Bond, Community and Organizational Care, and Shared Identity. Understanding these helps us unpack the construct of relatedness, identify points of intervention in a healthcare workplace context, and thus make the construct more tractable for HCI.

### 5.1 Connection of HCPs' relatedness layers with prior theories and empirical evidence

Several elements of each layer resonate with existing theories and evidence. For example, the second layer (Camaraderie and Bond) echoes prior evidence showing that deepened ties and solidarity play an indispensable role in HCPs' wellbeing and work engagement [2, 59, 61]. Forming deeper, trusting relationships with respect and concern for colleagues is also identified as a key relatedness satisfying element in the SDT literature on workplace motivation [30]. The third layer (Community and Organizational Care) aligns with the implications of the Belongingness Hypothesis [5] and Perceived Organizational Support (POS) [55]. These theories revealed insights into the need to connect with larger groups and organizations, and the importance of organizations valuing employees' contributions and caring about their wellbeing [55]. The final layer (Shared Identity) also intersects with constructs like Social Identity Theory, which asserts that people tend to categorize themselves into social groups, shaping their self-concept and identity [46]. Meanwhile, previous studies have also shown that HCPs and medical students widely endorse a strong professional identity [12, 65] and a shared "sense of calling" [93].

Although the layers themselves share conceptual overlaps with prior constructs, the underlying themes and qualitative data reveal fine-grained nuances in how relatedness is experienced by HCPs within their workplace. For example, our analysis showed that daily interactions need to be brief, informal, self-initiated, or autonomy-supportive to be beneficial. While other studies have highlighted the importance of social interactions for HCPs [60], the informal and autonomy-supportive nature of these interactions has been less emphasized. By underlining these qualities, we extend the mental health and HCI research on HCPs' workplace social support, suggesting both digital and non-digital interventions should prioritize user-led, autonomy-supportive forms of connection rather than prescriptive or coercive forms. Other aspects revealed in this study, like the value of personal support networks within the team, the relational benefits of organization-wide events, and how common memories and broader social-political contexts feed into HCPs' shared identity, are also less addressed in prior literature. We suggest that these aspects can be central design considerations for interventions and support on HCPs' relatedness, as they are grounded in HCPs' real-life experience and the unique contexts they work in.

Furthermore, our work seeks to provide a more holistic understanding of how relatedness is experienced by HCPs, offering a deeper account of the often-generic concept of "relatedness needs". This could complement existing SDT work by unpacking relatedness as a complex, multi-layered need. It is also essential to view the layers as interdependent rather than separate. For example, informal and everyday interactions (the first layer) among HCPs are highly individual and are hard to program directly. However, by acting on other layers, such as investing in shared social space and staff-oriented events (third layer), and emphasizing psychologically safe environments within teams (second layer), it may be possible to create time, space, and opportunities for increased interactions among HCPs. Previous studies have documented evidence for similar practices and their relational and wellbeing benefits [83, 96]. Similarly, HCPs' shared identity (fourth layer) can be enhanced by everyday collegial support, team solidarity, and visible organizational care that consolidates their feeling of togetherness. And a salient shared identity among HCPs, as implied by prior studies [52, 66], can in turn facilitate their initiating interactions with colleagues, offering help, and engaging in organizational and community initiatives.

**Table 2: Concept or categories of design ideas generated by HCPs**

| Concept (Category of Design ideas) | Concept/Category description | Relatedness support Layer(s) it pertains to | Example of design ideas in this category |
|---|---|---|---|
| Playful Encounter | Technology that facilitates ice-breaking or positive and playful brief interactions among colleagues. | Informal Interaction | A physical lamp on the desk that senses people sitting around it and provide feedback or stimulations for small talk, conversations or shared laughter (P12) |
| Collocated Action | Technology that help HCPs engage into collective experiences and shared activities. | Informal Interaction + Camaraderie and Bond | An on-screen AI-based Yoga coach in the communal garden or studio, that instructs staff members to do yoga or relaxation movement. (P9) |
| | | | Digital retreat in the staff room using VR or projection mapping, where HCPs can enjoy brief disengage from medical environment together. (P5, P12 and P16) |
| Effective Collaboration | Technology that support and enhance the teamwork and team cohesion within wards and units. | Informal Interaction + Camaraderie and Bond | Updated EHR system where HCPs can see AI's suggestions and similar cases of other HCPs and allow reciprocal support if they would like to. (P2) |
| Team-up Activities | Technology that allows HCPs engage as teams, creating opportunities for team building and interactions beyond their own circles. | Informal Interaction + Camaraderie and Bond + Community and Organizational Care | A challenge-based wellbeing game that allows users to log in either as individuals or as a team/ward. So that teams can bond with the playful experience and different teams can challenge each other in a playful context (P21 and P4). |
| Asynchronized Sharing and Support | Technology that encourages HCPs to share their thoughts and feelings in an asynchronized and discrete way, and to care and support others. | Camaraderie and Bond + Community and Organizational Care | A digital platform or a physical installation on which staff can post about their support need and others could leave comments and offer support accordingly. (P3, P4, P11) |
| Embodiment of Care | Technology that embodies and showcase the community care and resources available. | Community and Organizational Care | Smart digital signage that provide personalized suggestions and recommendations of the events, resources and staff wellbeing programs available in the hospital (P6). |
| Expressing Appreciation | Technology that can be utilized to express appreciation and gratitude towards colleagues, in the form of digital or tangible awards or gifts. | Camaraderie and Bond + Community and Organizational Care + Shared Identity | A digital wall of fame, for the appreciation of colleagues. Staff can post on the wall and give out virtual gifts for their appreciation for the colleagues who had offered support to them. (P14) |
| Memories and Stories | Technology that records and crystalize the moments, milestones, stories, and memories among HCPs. | Community and Organizational Care + Shared Identity | A selfie installation within the staff room that record key moments shared by colleagues, or just the day-to-day lives within the team. (P3 and P5) |

According to SDT, relatedness is one of the fundamental psychological needs, along with autonomy and competence [26]. However, in occupational motivation studies, competence and autonomy were usually given greater focus [26, 30, 56], while relatedness need is often simply defined as good inter-collegial relationships

[84]. Based on our findings, we argue that relatedness needs are complex and layered, and their satisfaction could benefit the autonomy and competence of HCPs. For example, a strong support network, team cohesion, and organizational or community connections could enhance HCPs' competence and make them more resilient at work, as shown by previous studies [52, 59]. Furthermore, HCPs with stronger, more internalized identities are more likely to be autonomously motivated in their daily work, as suggested by SDT literature [65]. Thus, through this work, we hope to highlight the importance of relatedness and to contribute to understanding the relatedness needs among HCPs.

## 5.2 Leveraging technologies for relatedness improvement for HCPs

As indicated by the concepts generated in our workshops, technology could enhance relatedness satisfaction across the four layers. Below we provide design considerations and actionable implementations on how technology could support relatedness among HCPs, within the unique healthcare context.

*5.2.1 Support brief, collocated connection that fits into HCPs' routine.* Many concepts generated in this study highlighted the importance of facilitating interactions, mutual care, and shared experiences among HCPs. The Playful Encounter, Collocated Action, and Team-up Activities concepts could support various relatedness layers, providing opportunities for direct interactions and deeper bonding within dyads or teams. This echoes similar themes from previous HCI literature. For example, Wenhart et al. highlighted recurrent strategies in relatedness technologies, like Joint action, Genuine Conversation, and Ritual Formation for strengthening connections in both close relationships and stranger encounters [90]. Other HCI projects also demonstrated the potential to satisfy relatedness needs through communal tangible games [31], computer-mediated self-disclosure [21], group-based exergames [25], and digital mindfulness sessions [40, 64].

However, it is worth noting that the contexts can vary significantly across subjects (e.g., general population vs. HCPs), and settings (office vs. clinical). Many insights from prior HCI works will need careful validation with HCPs, and within real-life healthcare environments. In particular, as discussed in Section 2, HCPs' heavy workload, complex work, unpredictable schedules, and focus on patient care could influence the appropriateness of technology interventions for social interaction. Previous studies on digital wellbeing interventions for HCPs have revealed that a lack of time and difficulty incorporating the interventions into daily routines are among the key barriers to engagement [45, 69]. Similarly, relatedness technologies need to account for these factors. For example, gamified relatedness technologies, as explored in [18, 31, 33], would require careful design to avoid overcomplicating HCPs' daily work, interrupting their schedules, or generating negative patient perceptions. Similarly, technologies for collocated interactions must be redesigned in terms of time requirements, cognitive load, and physical effort to fit the high-paced, safety-first nature of clinical practice. Similar concerns are reflected in a recent SDT-based study, revealing that high levels of relatedness satisfaction among nurses were associated with higher emotional exhaustion [8]. The authors further implied high relatedness satisfactions could became "too much of a good thing", especially for early-career nurses, as they may need to spend more time and effort developing and managing workplace relationships within an already demanding and time-intensive workplace, leaving them less time for their own wellbeing [8].

Taken together, we suggest that technologies facilitating HCPs' interactions be designed as workflow-aware, brief, and minimally disruptive. More importantly, relatedness technologies need to be designed to lighten HCPs' relational load, rather than adding more burden. We recommend designing concepts like Playful Encounter (e.g., social nudges or hints via screen-based or tangible interfaces) as brief, playful micro-rituals that can be incorporated into existing routines (e.g., after handovers, debriefs, or during lunch breaks, in places like staff rooms or canteens). Collocated Action and Team-up Activities (e.g., short collective exergames or digital-based group activities), which may require more time and effort, can be embedded around existing morning debriefs or between-shift breaks to prevent workflow interruptions. More importantly, such interventions should encourage but not enforce engagement, avoid creating intrusion or embarrassment for non-participation, and allow for flexible, come-and-go engagement.

*5.2.2 Deepen HCPs' connection while staying mindful of the hierarchy, professional boundaries, and cultural elements.* Furthermore, our results also demonstrated the importance of facilitating mutual care, appreciation, and recognition towards HCPs. The concepts like Expressing Appreciation, Asynchronized Sharing and Support and Embodiment of Care showcased HCPs' value for mutual care and support, as well as recognition and appreciation. This is consistent with existing evidence, which suggests that appreciation, recognition, and gratitude among colleagues are associated with greater team bonding and employee wellbeing [50, 54, 80]. In HCI, certain approaches have been explored, such as gamified employee recognition [70] (virtual gifts or badges among colleagues), tools that facilitate self-disclosure and conversations [21, 32, 43], and technologies that support acts of care [90].

However, in addition to the high workload and complexity, HCPs' work relationships are shaped by unique medical and organizational contexts, including professional hierarchies and boundaries [53, 72], close interdisciplinary collaborations [72], and cultural aspects such as stigma and professional norms [19, 29]. In such contexts, technologies that facilitate deeper conversation, self-disclosure, and mutual support can become inappropriate if not carefully designed. HCPs may be reluctant to disclose vulnerabilities to colleagues outside their usual support circle, particularly in hierarchical teams and departments. This reluctance is likely to increase if their communications feel surveilled or are directed towards their managers. It is understandable that concerns about privacy, stigma, limited trust, and professional norms are some main hindrances to HCPs' engagement with digital wellbeing tools, as identified by prior studies [45, 95].

Thus, we recommend developing relatedness technologies while staying mindful of the professional boundaries and hierarchies. Technology designed to support mutual caring and appreciation could incorporate features that let HCPs control the audience of their interactions (e.g., peer-only, team-only, or organization-wide). We also recommend future designs to clarify that any self-disclosure and expressions of support or appreciation are separated from performance management and surveillance. With these considerations, design concepts like one-click "appreciation", thankful notes tied to a handover/task via staff systems, and even a team- or organization-level virtual gratification wall where staff could provide support or honor their shared memories, could be implemented in a safe, trusted, and non-intrusive way to support staff connection and mutual respect.

*5.2.3 Build on existing digital ecosystems to avoid adding complexities.* Besides introducing new technologies, research could also explore how to enhance the current digital ecosystem within healthcare organizations to provide additional cues that support relatedness. This would allow more interactions, collaboration, and support to be embedded into HCPs' current routine, with fewer complexities incurred in their work. For example, integrating the concepts of Effective Collaboration, Asynchronized Sharing and Support, and Embodiment of Care into staff-facing apps or EHR systems could support increased interaction, bonds, and a sense of community. Recent studies have demonstrated that adding social functions like secure chat [4, 15], and asynchronous messaging (e.g., a microblog-style messaging system [24]), to EHR systems has improved cross-disciplinary collaboration, team cohesion, and support. Robertson et al.'s review further suggested that EHRs, which have repeatedly been shown to reduce social connectedness, could be proactively designed to restore workplace relatedness [71].

Given the well-documented time pressure and digital overload among HCPs [79, 95], we argue that future interventions should consider starting from the existing workflows and digital infrastructures rather than jumping straight into designing a new digital intervention. Future work could transfer existing HCI insights on relatedness technologies, such as increasing awareness of peers [43, 90] and supporting expressivity and mutual care [21, 43], to healthcare digital ecosystems, such as EHRs and staff apps. For example, ambient cues to raise awareness of colleagues' status and availability for interactions [77, 91] might be redesigned and integrated into existing EHR systems. Similarly, an asynchronous version of a peer-support program, such as "Schwartz Rounds" [35], could be incorporated into staff-facing apps to facilitate smoother user adoption and flexible engagement.

To summarize, although previous HCI works on relatedness technologies have demonstrated many insights, greater attention should be taken when implementing HCP-generated concepts into real life. We therefore recommend future works to adopt a participatory approach, by co-designing relatedness interventions with stakeholders including HCPs, healthcare managers, and even patients and policymakers, to fully understand the users (HCPs) and their contextual constraints. Meanwhile, as suggested by the NASSS (non-adoption, abandonment, scale-up, spread, and sustainability) framework [36], we argue that designers and researchers should not treat relatedness technologies as plug-and-play solutions, but instead as part of the digital infrastructure that aligns with staff needs, organizational capacity, system requirements, funding and regulatory contexts, and long-term impacts. This could help future relatedness and wellbeing technologies move beyond isolated prototypes and generate sustainable benefits for HCPs.

### 5.3 Future directions: HCP relatedness amid the evolution of healthcare systems

Besides design implications, the layered model for relatedness needs can also be relevant to broader systematic shifts in healthcare. Growing digitalization in healthcare systems around the world has been documented especially since COVID-19 [79]. In the UK, the recent NHS 10-Year Plan [27] demonstrated a shift toward a "digital by default" approach to future healthcare, despite findings that some digital platforms, such as the EHR, can have detrimental effects on HCPs' relatedness and social connection [71]. This intersects with another shift, which is the reorganization of clinical work into smaller, hybrid, and distributed teams. The UK has unveiled its plan to shift healthcare from centralized hospitals to distributed community-based teams [27]. Similar situations can be seen in the US and Europe, as well as in countries in the Global South [97]. Such a shift could further challenge the existing human connections among HCPs. Our participants (like D04) recognized this change, noting a decline in the "feeling of a team" as more colleagues began hybrid working after COVID-19.

Luckily, technologies can also create opportunities. As HCPs become more proficient with digital systems (potentially with the help of AI), the time saved could be redirected to restoring human connections. For example, AI systems such as Ambient Scribe Tools that record conversations between clinicians and patients, and subsequently assist in writing patient notes, have been implemented in hospitals in the US [28]. Similarly, in the UK, the first-ever NHS patient was discharged with AI-based notes in 2025. It is reasonable to expect broader deployment of similar systems to relieve administrative formalities and after-hours charting in the foreseeable future. However, AI tools will *not* automatically repair lost connections among HCPs. We advocate that any time and cognitive load relieved by AI should be protected by institutions so it can be responsibly 'reinvested' into strengthening positive human interaction, bonds, sense of community, and shared identity. Hence, there is an urgent need for designers and developers to work with organizations and policymakers to embed relatedness-supportive principles into rapidly evolving AI-enabled healthcare systems. We hope the work described herein takes some first steps towards informing such a shift.

### 5.4 Limitations

This study has some important limitations. Firstly, although we speak broadly of health care professionals, the study was carried out in a single hospital in the UK, with a volunteer sample of HCPs. While some of the relatedness challenges faced by HCPs are likely to be common across cultures and healthcare systems [61], our findings may not transfer directly to other sites and countries. Future studies should examine how the need for relatedness among HCPs differs across cultures and institutions. In addition, the volunteer sampling method and the reliance on self-reported data may have introduced self-selection, and self-report biases, including social desirability and retrospective sense-making. Future research should consider involve more diverse participant groups and use randomized recruitment methods to reduce the risk of self-selection bias. In addition, the interview guide inclined towards stress and stress management, which may have influenced participants' reflections and feedback.

Meanwhile, this study is mostly qualitative and exploratory, with a limited dataset. The four-layer model of relatedness we identified should be seen as an exploratory framework to guide further work. More studies are needed to evaluate whether the current model can be transferred into quantitative measures. Confirmatory studies are needed to evaluate the causal impact of interventions based on this model of HCP relatedness.

Furthermore, this study only focused on workplace relationships among colleagues, rather than HCP-patient relationships or their own personal relationships. Although HCP-patient connections are an important source of relatedness for clinicians [2], they are fundamentally different, shaped by professional responsibility, altruism, power asymmetries, and professional ethics and confidentiality. Including both types of relationships in one study may weaken the specificity of our findings and design implications.

Finally, our co-design workshops focused on low-fidelity idea generation. Despite efforts to mitigate the possibility, there still could be partial overlap between interview and co-design participants, which may have reinforced certain perspectives across data sources. And at the current stage of the study, we were unable to fully test developed prototypes in organic hospital settings. Future studies could move towards pilot implementations of technologies embedded in everyday healthcare digital ecosystems.

# 6 CONCLUSIONS

In this study, we conducted semi-structured interviews and co-design workshops with HCPs to explore how their need for relatedness is supported at work, and how technologies could be designed to facilitate relatedness need satisfaction. Interview data informed a four-layer model of relatedness needs among HCPs, which highlighted the significance of informal interactions, camaraderie, community support, and shared identity for HCPs' sense of relatedness. Building on the workshop data, we further distilled several design concepts (e.g., Playful Encounters, Collocated Action, Memories & Stories) that categorize participant-generated ideas into actionable directions for HCI. Finally, we compared our findings with existing psychological theories, including SDT, and empirical evidence both in psychology and HCI, and discussed how the exploratory model and design implications can be beneficial in the unique healthcare context. As healthcare becomes more digitalized and AI-enabled, we argue that supporting HCPs' relatedness should be treated as an urgent design priority. We hope this study could inform and motivate future research on sustaining meaningful connections among HCPs.

## Acknowledgments

This project is supported by CW+ and the Chinese Scholarship Council. We would like to thank all the participants who were involved in the interviews and workshops.

## APPENDICES

### Appendix A

Interview purposes:

- To explore how healthcare professionals (HCPs) are currently receiving and providing relatedness support, as well as their perceptions of the relatedness-support strategies and methods.
- **(Sub-goal)** To understand HCPs' present working contexts and identify how relatedness support is (or could be) integrated.

Interview questions:

1. Daily routine

   1) How often/how long do you take breaks daily?
   2) What will you usually do to relax/take a break from work?
   3) Do you notice any chances for relatedness support in your daily routine?

2. Work-related Stress and Burnout

   1) Do you often find yourself stressed by work? What are the main sources of this work-related stress?
   2) How do you usually cope with work-related stress? Is there something you do/think of to relax when you are feeling stressed?
   3) Would you seek help from colleagues when experiencing stress and burnout?
   4) How do you think social support, or a sense of relatedness, has helped you reduce or cope with stress?

3. Relatedness support methods

   1) What are the current opportunities provided by your employer (the hospital) to reduce stress and burnout? Is there anything related to support for relatedness?
   2) (If there is,) Do you think the current support is helpful? Why?
   3) Have you tried to seek opportunities to enhance social support or connect with colleagues?
   4) (If not mentioned in the previous discussion,) Could you please elaborate on your thoughts on the strategies/programs on relatedness/social support:

a. Peer-support Groups;
b. Community Support:
c. Digital Social Support

### Appendix B

Below are the introductory materials provided to the participants in the workshops. These materials were shared with the participants in the form of both printed sheets (Figure 4) and digital presentation slides (Figure 5 and 6) at the beginning of each workshop.

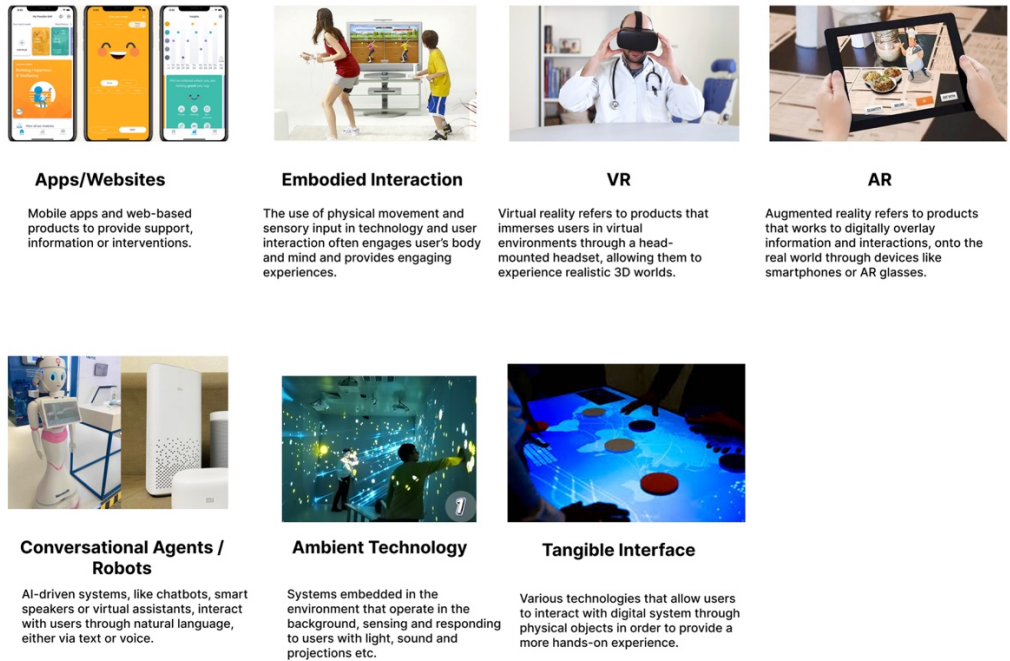

**Figure 4: Printed information on potential technologies could be used to support wellbeing and relatedness needs**

## Potential Technologies - Virtual Reality (VR)

**Definition**

**Case Studies**

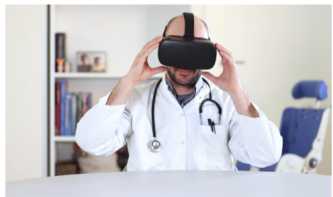

Virtual reality refers to products that immerses users in virtual environments through a head-mounted headset, allowing them to experience realistic 3D worlds.

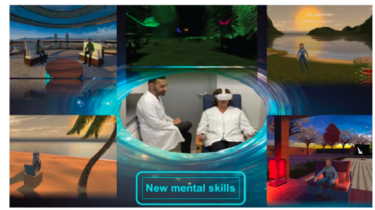

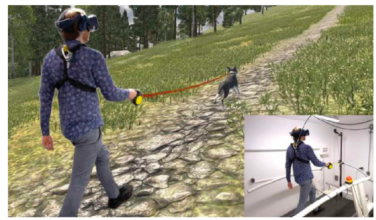

Ferrer Costa et al. [2] and Halliburton et al. [3]

**Figure 5: Example of the digital presentation slides on technologies and case studies**

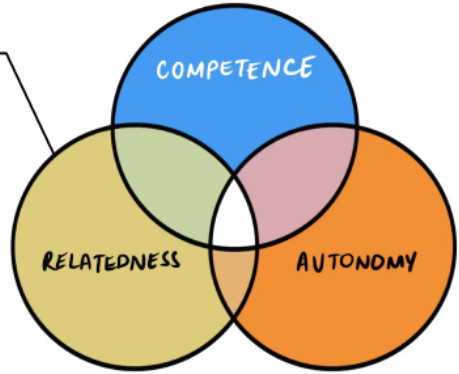

**Figure 6: Example of the digital presentation slides on Self-Determination Theory and relatedness**